\documentclass{PoS}

\title{Gravitational lensing at milliarcsecond angular resolution with VLBI observations}

\ShortTitle{Gravitational lensing at milliarcsecond angular resolution}

 \author{\speaker{C. Spingola}$^1$, J. P. McKean, $^{2}$, A. Deller$^{3}$, J. Moldon$^{4}$ \\
     \llap{$^1$}  Kapteyn Astronomical Institute, University of Groningen, Postbus 800, NL--9700 AV Groningen, the Netherlands \\
     \llap{$^2$} ASTRON, Netherlands Institute for Radio Astronomy, Oude Hoogeveensedijk 4, 7991 PD Dwingeloo, the Netherlands\\
     \llap{$^3$} Centre for Astrophysics and Supercomputing, Swinburne University of Technology, John St, Hawthorn, VIC 3122, Australia\\
      \llap{$^4$} Jodrell Bank Centre for Astrophysics, Alan Turing Building, The University of Manchester, Oxford Road, Manchester, M13 9PL, UK\\ 
    E-mail: \email{spingola@astro.rug.nl}\\
     } 

\abstract{Gravitational lensing is a powerful tool for quantifying the mass content and distribution in distant galaxies. By using milliarcsecond angular resolution observations of radio-loud gravitationally lensed sources it is also possible to detect and quantify small deviations from a smooth mass density distribution, which can be due to low mass substructures in the lensing galaxy. We present high-resolution global VLBI observations of the gravitationally lensed radio source MG J0751+2716 (at z = 3.2), that shows evidence of both compact and extended structure (core-jet morphology) across several gravitational arcs. These data provide a wealth of observational constraints that are used to determine the inner (baryonic and dark matter) mass profile of a group of galaxies and also investigate the smoothness of the dark matter distribution on mas-scales, which is sensitive to possible structures of $10^{6-7}$ M$_{\odot}$ within the lensing halo or along the line-of-sight. Our lens modelling finds evidence for astrometric anomalies in this system, which suggest presence of extra mass structure in the lens model. To date this kind of detailed studies of gravitational lensing systems like MG J0751+2716 has been limited by the currently small sample of radio-loud gravitational lenses. In this context, we also present a new pilot gravitational lens search in the VLBI survey \textsl{mJIVE--20}, in perspective of future surveys with the next generation of radio interferometers. }

\FullConference{14th European VLBI Network Symposium \& Users Meeting (EVN 2018)\\
		8-11 October 2018\\
		Granada, Spain}
\begin{document}

\section{Introduction}
According to the theory of general relativity, a gravitational field bends space-time and, therefore, the geodesic of a light ray is also bent because of the presence of a massive object. As a consequence, multiple distorted magnified images of a background source may be observed if a massive object (acting as a lens) lies between the distant source and the observer. Therefore, the observation of gravitationally lensed sources allows us to probe the total (baryonic and dark) matter density distribution in distant galaxies acting as a lens (see Treu 2010 for a review). Moreover, when gravitational lensing is combined with high angular resolution observations, it is then possible to access the crucial small scales needed to test the validity of the $\Lambda$CDM structure formation model (e.g. Bullock et al. 2017).

In order to accurately constrain the mass density distribution of lensing galaxies, systems with a radio-loud background source have some important advantages. First, the radio emission is not reddened nor obscured by the dust or the usually bright optical emission from the lensing galaxy (e.g. York et al. 2001). Therefore, the lensed images can be easily identified and their flux density can be confidently measured. Moreover, radio-loud sources often have extended structure, which can, for instance, be due to the presence of AGN jets, that extend from several pc- to kpc-scales. This means that gravitationally lensed radio sources can show extended multiple images, potentially connected into gravitational arcs, making them the most suitable systems for detailed studies of the mass density distribution of the foreground lensing galaxies (King et al. 1997, Biggs et al. 2004, More et al. 2009). 

As lens galaxies are typically massive ellipticals, with a mass of 10$^{11-12}$ M$_{\odot}$, the maximum image separations in most galaxy-scale systems is about 1.2 arcsec (Turner et al. 1984, Browne et al. 2003, Koopmans et al. 2009). Therefore, in order to spatially resolve the multiple lensed images, it is necessary to observe lensing systems at sub-arcsec resolution. Moreover, for a detailed study of the mass density distribution of the deflector, a precise astrometry of the lensed images, at possibly sub-mas levels, is required. Therefore, only mas angular resolution observations can investigate the small-scale properties of the mass density distribution of a lensing galaxy, and Very Long Baseline Interferometry (VLBI) can achieve such high angular resolution at cm-wavelengths. This high angular resolution is needed, for instance, to investigate gravitational lensing by sub-haloes at the $10^6$ M$_{\odot}$ level and test the smoothness of galaxy-scale dark matter haloes on small scales (e.g. Li et al. 2016).

\section{Astrometric anomalies in the gravitational lensing system MG~J0751+2716}

\begin{figure}
    \centering
    \includegraphics[width=0.7\textwidth]{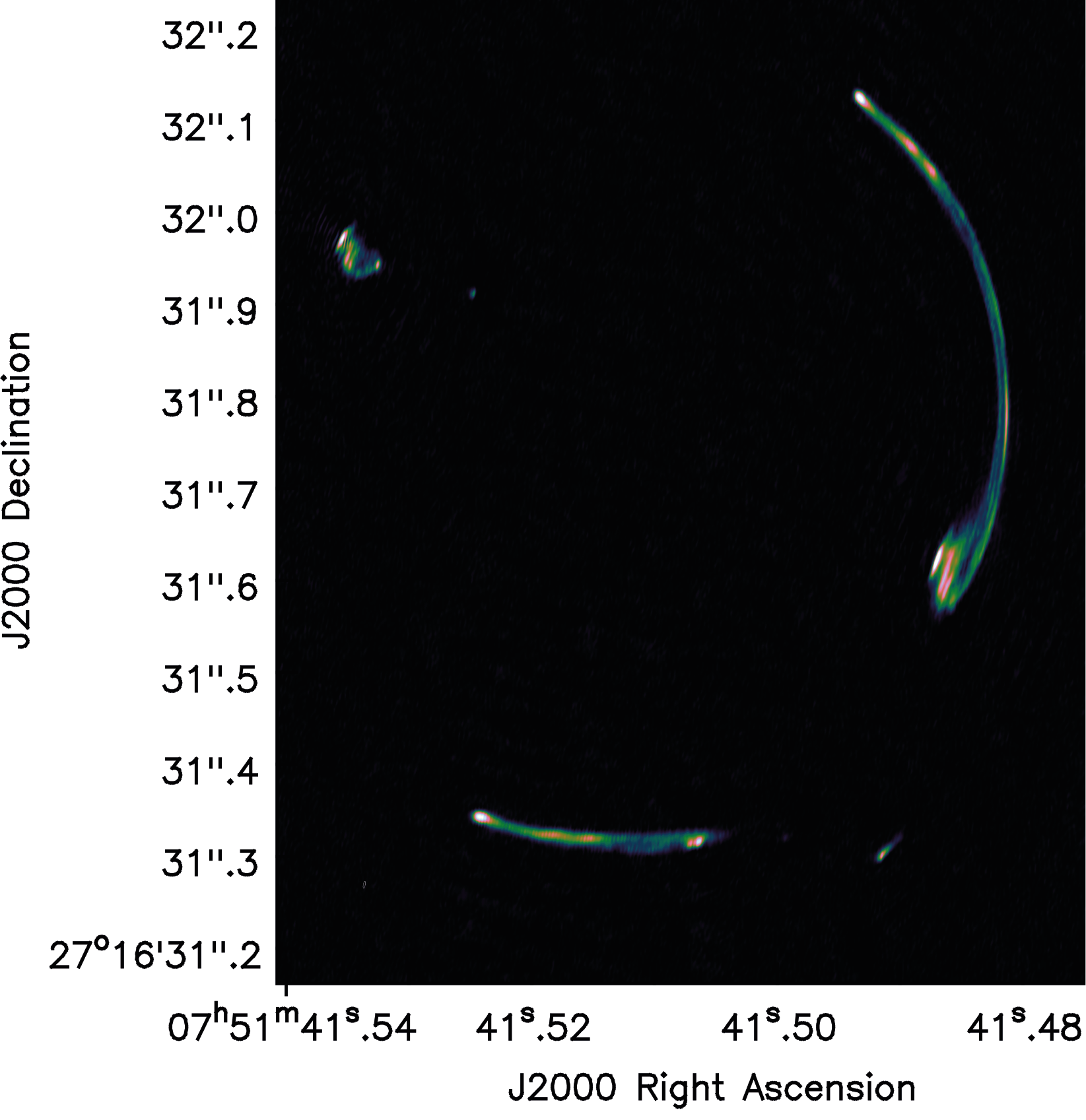}
    \caption{Global VLBI imaging of the radio-loud lensing system MG~J0751+2716 at 1.65 GHz, using uniform weights and multi-scale cleaning within {\sc wsclean}.}
    \label{fig:1}
\end{figure}

\begin{figure}
    \centering
    \includegraphics[width=\textwidth]{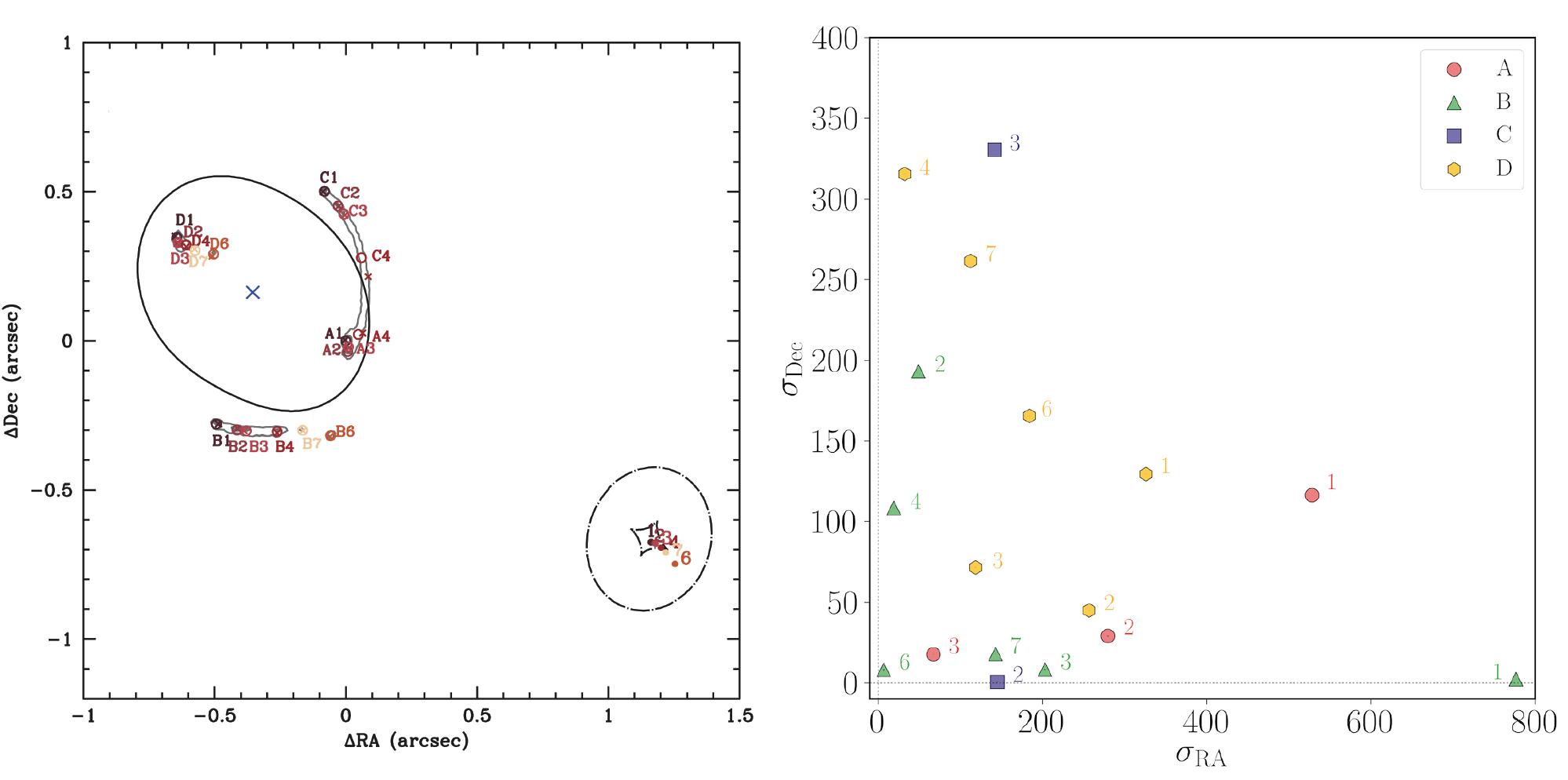}
    \caption{(Left) Schematic representation of the more realistic lens mass model for MG J0751+2716 described in Sec. 2. 
    The position of the main lensing galaxy is indicated by the blue cross.  The observed positions are the open circles and the predicted positions are represented by the crosses, with all positions given relative to component A1. Each colour corresponds to a different background source component (filled circles). The lens critical curve is shown by the thick line; the dashed line shows the source plane caustics. The grey lines are the $3\sigma$ contours of the extended emission detected from our global VLBI imaging (shown in Fig. \ref{fig:1}), for reference.
    (Right) The offset between the observed and the model-predicted positions in units of sigma for the lens model shown in the left panel. Each colour and symbol represents a different group of lensed images. The error bars are shown in black and the two black dashed lines indicate the no offset position.}
    \label{fig:2}
\end{figure}

 The background jetted active galactic nucleus (AGN) at redshift z = 3.2 is gravitationally lensed by a foreground group of galaxies at redshift z = 0.35 (Lehar et al. 1997, Tonry et al. 1998, Momcheva et al. 2006). The distorted view of the distant radio jets consists of four spectacular gravitational arcs that are extended by 200 to 600 mas; these observations represent the highest angular resolution imaging of extended gravitational arcs produced by a gravitational lens (Fig. \ref{fig:1}). The mas-resolution global VLBI observations of these gravitational arcs provide a large number of constraints to the lens mass model, which are used to search for possible deviations from a globally smooth mass distribution on small angular scales (Spingola et al. 2018a). By identifying groups of lensed images that corresponds to the same part of the background source, we investigate two models for the MG J0751+2716 lensing mass density distribution.  One model includes the main lensing galaxy only plus an external shear term, while the other (more realistic model) includes all the haloes part of the group of galaxies and an external shear component. The latter is shown in Fig. \ref{fig:2}.

 Given the exquisite quality of the radio imaging, the lens mass model parameters are inferred with a precision of less than a per cent, even though the simple smooth models are not accurate enough to fit the positions of the observed images to within the measurement error level. 
 Furthermore, the slope of the mass density distribution of the main lensing galaxy is steeper than isothermal at the 4.2$\sigma$ level and at the 6.8$\sigma$ level for the two models, respectively. This mass density slope is consistent with studies of low-mass early-type satellite galaxies within groups or clusters (e.g. Auger et al. 2008, Tortora et al. 2014), and can be taken as evidence that supports the two-phase galaxy formation scenario (Guo \& White 2008).
 Therefore, gravitational lensing can provide a description of the projected mass density distribution of complex galaxy groups, which is generally challenging to trace at high redshift. 
 
 Moreover, the excellent sensitivity and high angular resolution of the VLBI imaging of MG J0751+2716 allowed the measurement of small offsets between the observed and predicted positions of the lensed images, with an average positional rms of the order of 3 mas for our simple parametric models (Fig. \ref{fig:2}). These astrometric offsets suggests that, at mas level, the standard assumption of a smooth mass distribution fails, requiring additional mass structure in the model. Therefore, this result challenges the hypothesis that galaxy haloes are smoother than what predicted by the $\Lambda$CDM model to explain the low number of sub-structures associated with massive haloes (e.g. Lovell et al. 2012). However, this extra mass cannot be unequivocally attributed to a population of low mass sub-haloes within the lensing galaxy, as predicted by cold dark matter models for galaxy formation, but may be due to a more complex mass distribution within the lensing group or along the line-of-sight (Despali et al. 2018). A source model that takes into account the entire flux density of the gravitational arcs is necessary to test our current lens models. By using grid-based corrections to the gravitational potential, based on the methodology of Vegetti \& Koopmans (2009), it will be possible to infer the origin of the astrometric anomalies in MG J0751+2716.
 
\section{Gravitational lens search in the mJIVE--20 survey}

The next generation of radio survey facilities, such as the Square Kilometre Array (SKA), will allow the discovery of $\sim10^5$ strong gravitational lenses, increasing the number of radio-loud lensing systems by four orders of magnitude (Koopmans et al. 2004, McKean et al. 2015). These surveys will also provide a large parent sample for detailed studies of individual objects (with SKA-VLBI, for example, Paragi et al. 2015), as described in the previous section for the gravitational lensing system MG J0751+2716.
What strategy should we follow in order to efficiently find $\sim10^5$ radio-loud lensed sources? 

\begin{figure}
    \centering
    \includegraphics[width = \textwidth]{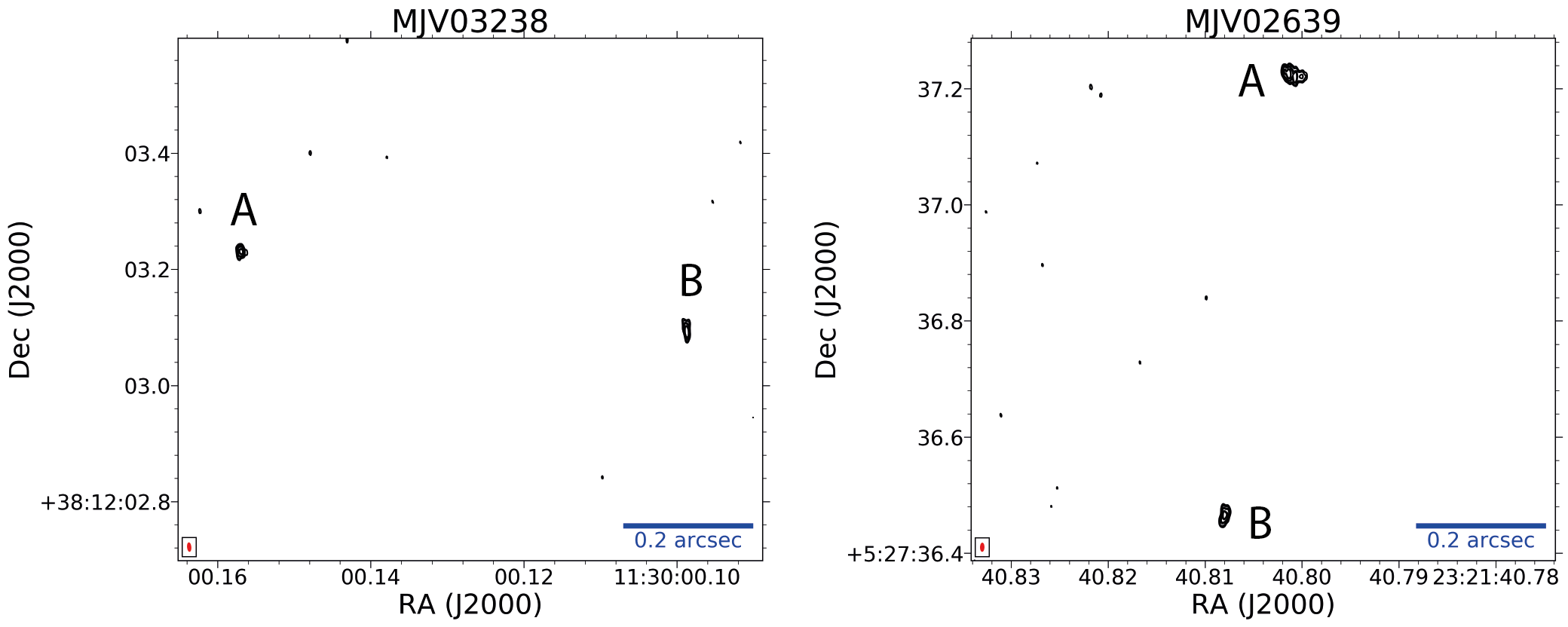}
\includegraphics[scale = 0.28]{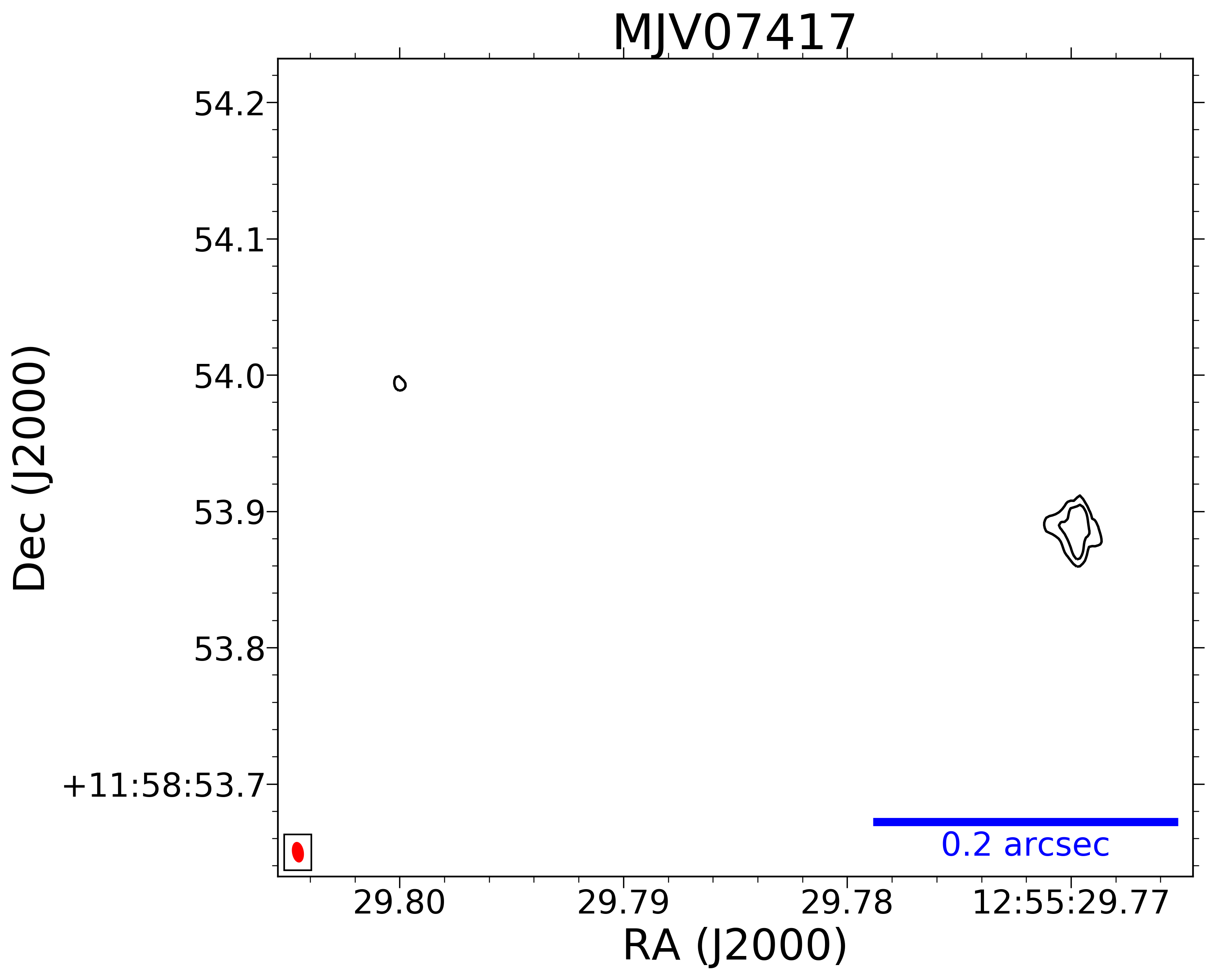}
    \caption{(Upper-panel) Self-calibrated images at 1.4 GHz of two mJIVE--20 lens candidate that are known lensing systems discovered by CLASS (B1127+385 on the left and B2319+051 on the right). Contours are at $(-3, 3, 6, 9, 12, 15, 27) \times \sigma_{\rm rms}$, the off-source rms noise. The beam size is shown in the bottom left corner, which is $11 \times 5$ mas$^2$; north is up and east is left. The blue scale bar in each image represents 0.2 arcsec.
    (Lower-panel) Cleaned image at 1.4 GHz of the final lens candidate MJV07417 from the mJIVE--20 samples, same legend. }
    \label{fig:3}
\end{figure}

We performed a pilot search for gravitational lensing systems in the 
mJy Imaging VLBA Exploration at 20 cm survey (mJIVE--20, Deller and Middelberg 2014), which observed 24\,903 radio sources selected from FIRST with the VLBA at an angular resolution of 5 mas. 
Following a strategy similar to Cosmic Lens All Sky Survey (CLASS, Myers et al. 2003, Browne et al. 2003), we have inspected an initial sample 3\,640 snapshot observations, with the aim of searching multiple components that can be associated with a gravitational lensing system (Spingola et al. 2018b). We selected as best lens candidates the sources that showed multiple components separated by $>$ 100 mas, with a flux-ratio of $\leq15$ and a surface brightness consistent with gravitational lensing. Among the candidates, two consists of re-discoveries of known gravitational lenses, which were part of the CLASS survey. These two systems are B1127+385 and B2319+051, and they are shown in Fig. \ref{fig:3} (Koopmans et al. 1999, Rusin et al. 2001). The remaining twelve candidates were then re-observed with a longer exposure time at 1.4 GHz, and then followed-up at higher angular resolution at 4.1 and 7.1 GHz with the Very Long Baseline Array (VLBA). This step was fundamental to measure the spectral index and surface brightness of the individual components as a function of frequency. Ten of the candidates were rejected as the high angular resolution imaging revealed AGN core-jet or core-hotspot(s) systems, with surface brightness distributions and/or spectral indices not consistent with being gravitationally lensed sources, and the rejection of one system
was based on the lens modelling. The final lens candidate is shown in Fig. \ref{fig:3}, and has an image configuration that is consistent with a simple lens mass model, although further optical imaging and spectroscopic observations are required to confirm its possible lensing nature. 

Given the two known gravitational lensing systems, we estimate a robust lensing rate of 1:(318 $\pm$ 225) for a statistical sample of 635 radio sources detected on mas-scales, which is consistent with that found for CLASS within the (large) uncertainties (Browne et al. 2003).

\section{Summary}

Observing distant radio-jets in a lensing configuration that can produce extended gravitational arcs is rare. However, even a single lensing system that shows extended structure on milliarcsecond scales can provide a wealth of information on the mass distribution of the lensing galaxy at the smallest angular-scales, which otherwise would not be possible to access at high redshift.

Thanks to the excellent sensitivity and angular resolution of global VLBI observations combined with gravitational lensing, it was possible to reveal the high complexity on mas-scales of the mass the mass distribution in a lensing galaxy at redshift $z=0.35$. The observed astrometric anomalies in MG~J0751+2716 challenge the simple assumption that the mass density profile of a lensing galaxy can be considered smooth at mas resolution. However, in order to statistically constrain the mass distribution of lensing galaxies on sub-kpc scales we need a larger sample of lensing systems like MG~J0751+2716. In this respect, we have carried out the first wide-field VLBI lens search in the mJIVE--20 survey.
Following a methodology analogous to the CLASS lensing surveys, we showed that compact radio sources that are gravitationally lensed can be easily identified from their radio structures, when multi-frequency information is available. However, the identification of those rare cases with extended gravitational arcs, like MG~J0751+2716, may require a more sophisticated technique in order to not be contaminated by, for example, extended AGN jets. While this search has been limited by the $uv$-coverage of snapshot VLBI observations, in the future the large number of SKA antennas will provide a better imaging fidelity, and will reach a sensitivity of a few $\mu$Jy beam in minutes. This properties will make possible the detection of low surface brightness radio-loud compact sources and those with extended gravitational arcs. Moreover, if the SKA will have a VLBI capability, it will be possible to probe even smaller angular-scales of the lensing galaxy mass profile by following-up at higher angular resolution the most promising systems, as MG~J0751+2716. A large number of gravitational arcs emitting at the radio wavelenghts on VLBI-scales with robust lens models can allow the detection of low mass sub-haloes (as dwarf galaxies), whose abundance is a direct test for the $\Lambda$CDM galaxy formation model.

\end{document}